\documentclass[aps,pra,twocolumn,showpacs,floatfix,epsfig,epsf]{revtex4}
\usepackage{amsmath}
\usepackage{amssymb}
\usepackage{graphicx}
\usepackage{subfigure}
\usepackage{natbib}
\usepackage{color}
\usepackage{epstopdf}
\usepackage{keyval}
\usepackage{trig}
\usepackage{revsymb}

\begin{document}

\title{Interacting double dark resonances in a hot atomic vapor of helium}

\author{S. Kumar$^1$}
\author{T. Laupr\^etre$^{2}$}%
\author{R. Ghosh$^1$}%
\email{rghosh.jnu@gmail.com}
\author{F. Bretenaker$^2$}
\author{F. Goldfarb$^2$}
\affiliation{$^1$School of Physical Sciences, Jawaharlal Nehru University, New Delhi 110067, India\\
$^2$Laboratoire Aim\'e Cotton, CNRS - Universit\'e Paris Sud 11, 91405 Orsay Cedex, France}

\date{8 July 2011}

\begin{abstract}
We experimentally and theoretically study two different tripod configurations using metastable helium ($^4$He*), with the probe field polarization perpendicular and parallel to the quantization axis, defined by an applied weak magnetic field. In the first case, the two dark resonances interact incoherently and merge together into a single EIT peak with increasing coupling power. In the second case, we observe destructive interference between the two dark resonances inducing an extra absorption peak at the line center.
\end{abstract}


\maketitle

\section{Introduction}
Electromagnetically induced transparency (EIT) in three-level $\Lambda$-systems is a phenomenon in which an initially absorbing medium is rendered transparent to a resonant weak probe laser when a strong coupling laser is applied to a second transition \cite{Harris97}. In addition to its being a quantum interference phenomenon of fundamental interest, EIT has been studied extensively for its numerous applications, such as in slow and fast light, light storage, sensitive magnetometry and optical information processing. An extension of the usual three-level EIT to a four-level double-EIT scheme has been shown to expand the utility of EIT in a number of additional, potentially useful and easily controllable coherent nonlinear effects. These include engineering atomic response by perturbing a dark state \cite{Lukin99,Yelin03,Liang05}, and control of group velocity via interacting dark resonances \cite{Mahmoudi08}. The interest in four-level tripod-like atomic configurations started with an early work \cite{Bergmann96} showing that in a tripod medium, there exists an internal state subspace spanned by two orthogonal dark states, which is immune to spontaneous decay. Subsequently, the creation and measurement of a superposition of quantum states using stimulated Raman adiabatic passage in a tripod have been proposed and demonstrated \cite{Bergmann03,Unanyan04}. Simultaneous enhancement and suppression of a dark resonance have been observed by nondegenerate four-wave mixing in a solid in a tripod-like level configuration \cite{Ham00}. Large cross-phase modulation induced by interacting dark resonances in a tripod system of cold $^{87}$Rb has also been reported \cite{Peng08}. 
Tripod configurations with two probes and a common coupling beam have been studied in a variety of contexts -- theoretically for the magneto-optical Stern-Gerlach effect \cite{Sun08}, for the experimental demonstration of light storage at dual frequencies \cite{Karpa08}, in a proposal for all-optical quantum computation with efficient cross-phase modulation, and sensitive optical magnetometry \cite{Petrosyan04}, and experimentally for matched slow pulses using double EIT in Rb \cite{Lvovsky08}, and also for the study of nonlinear Faraday effect in an inverted $Y$ model in Rb vapor \cite{Drampyan09}. Interacting dark resonances in tripod configurations with two coupling beams and a common probe have been used to obtain sub-Doppler and subnatural narrowing of an absorption line, theoretically by Goren {\it et al.} \cite{Goren04}, and experimentally by Gavra {\it et al.} in Rb \cite{Gavra07}. A similar scheme has been suggested for applications in logic gates and sensitive optical switches \cite{Zhang07}.

There are several such applications of controllable double dark resonances in four-level tripod systems, and there are not many experimental results on tripod systems reported in the literature so far. In this context, we wish to probe a simple system of $^4$He* at room temperature. This medium has been shown to be an ideal candidate for achieving ultra-narrow (less than 10 kHz) EIT in a three-level $\Lambda$-system involving only electronic spins in the presence of Doppler broadening \cite{Goldfarb08}. We have confirmed the true nature of the two-photon process of EIT by our observation of asymmetric Doppler-averaged Fano-like transmission profiles in the presence of single-photon optical detunings. $^4$He* has some peculiar favorable properties: (i) Velocity-changing collisions enable us to span the entire Doppler profile \cite{Joyee09}. (ii) The absence of nuclear spin simplifies the level scheme and eliminates the need for repumping lasers compensating for losses into the other ground state hyperfine levels. (iii) Diffusive motion increases the transit time of the atoms through the laser beam and hence the Raman coherence life-time. (iv) Collisions with the ground state atoms do not depolarize the colliding $^4$He*. Thus there are no background atoms to contribute to noise. (v) Penning ionization among identically polarized $^4$He* atoms is almost forbidden \cite{Shlyapnikov94}. In the present work, we show that $^4$He* is a suitable candidate for realizing a clean four-level tripod system in a room-temperature gas. The excited state 2$^3P_{0}$ ($m_{e}$ = 0, $\vert e \rangle$) of $^4$He* can be coupled selectively to the 2$^3S_{1}$ sublevels, $m_{g}$ = $-1$ ($\vert g_{-} \rangle$), 0 ($\vert g_{0} \rangle$) and +1 ($\vert g_{+} \rangle$), by co-propagating laser beams at around 1083 nm, with $\sigma^{+}$, $\pi$ and $\sigma^{-}$ polarizations, respectively. The energy separation between the 2$^3P_{0}$ and the next lower sublevel 2$^3P_{1}$ is large (29.6 GHz) compared to the Doppler width ($\approx$ 1 GHz), allowing one to ensure that each transition is isolated. This is not the case, for example, in Rb \cite{Gavra07}.

We focus on two different tripod configurations based on the interaction of the atoms with linearly polarized coupling and probe beams in the presence of a horizontal transverse magnetic field. In the first case, the probe beam has vertical linear (V) polarization and the coupling beam has horizontal linear (H) polarization, while in the second case, the probe beam has H-polarization and the coupling beam has V-polarization. We can easily switch from one configuration to the other by just a change in the orientation of a wave-plate used in the set-up. The difference in the configurations comes from the number of probe transitions used, yielding a distinctive interplay of double dark resonances in each case. With $^4$He* at room temperature, using a weak magnetic field and polarization selective transitions mentioned above, we experimentally realize a tripod configuration with two probed transitions in the first case, and observe that the double dark resonances add incoherently, as there is no coherence between the two populated probe ground levels. This configuration serves as a useful reference for the second configuration studied with two coupling transitions. In the second case, double dark resonances are related to coherent population trapping in the ground states. As the number of excited states is less than the number of ground states, transfer of coherence does not play any role \cite{Taichenachev99}. Thus these double dark resonances are not stimulated Raman peaks \cite{Goren04,Meshulam07} but detuned EIT peaks, interfering destructively with each other leading to an absorption dip in-between for non-zero magnetic fields. We model the system successfully and verify our experimental results.

The paper is organized as follows. In Sec.\,II, we describe the experimental set-up. In Sec.\,III, we present the experimental results and compare them with our numerical simulations for the two different tripod configurations. Our conclusions are presented in Sec.\,IV, with hints of potential applications.

\section{Experimental set-up}

\begin{figure}[htbp]
\begin{center}
 \scalebox{0.63}{\includegraphics{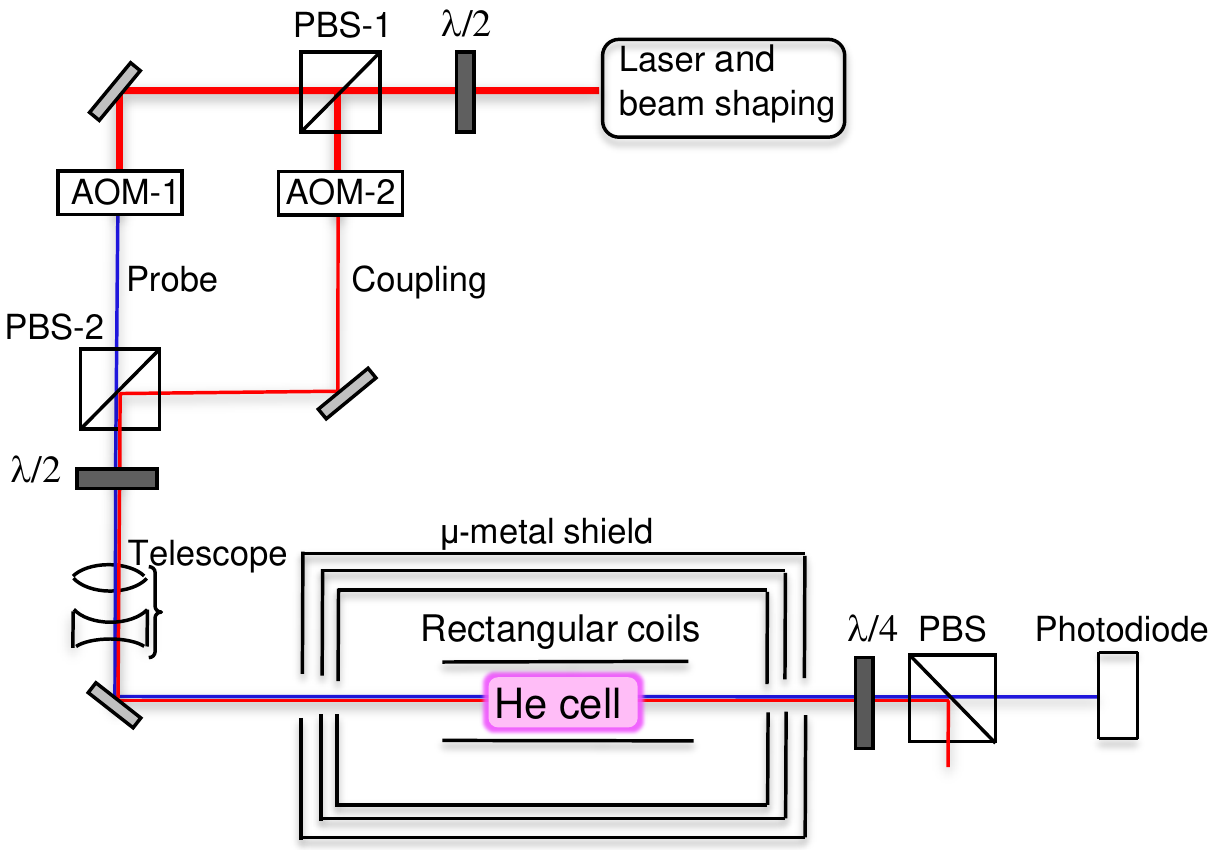}}
\end{center}
\caption{(Color online) Experimental set-up. PBS: polarizing beam-splitter, AOM: acousto-optic modulator, $\lambda/2$: half-wave plate, $\lambda/4$: quarter-wave plate}
\label{Expfig}
\end{figure}

The experimental set-up is shown in Fig.\,\ref{Expfig}. The helium cell is 6-cm long and has a diameter of 2.5 cm, and is filled with $^4$He at 1 Torr. The cell is placed in a three-layer $\mu$-metal shield to isolate the system from the earth's magnetic field inhomogeneities. Helium atoms are excited to the metastable state by an RF discharge at 27 MHz. We use the 2$^3\mathrm{S}_1 \rightarrow 2^3\mathrm{P}_0$ transition of $^4$He* ($D_{0}$ line) with the coupling and probe beams derived from a single laser at nearly 1082.9 nm wavelength (linewidth $\simeq$ 10 MHz), with a beam diameter of 1 cm after the telescope. The maximum available power for the coupling beam is about 27 mW, which is large enough due to the fact that the saturation intensity in $^4$He* is very low (0.167 mW/cm$^2$). A probe power of 100 $\mu$W has been used throughout. The frequencies and intensities of the coupling and probe beams are adjusted by the amplitudes and the frequencies of the RF signals driving the acousto-optic modulators AOM-1 and AOM-2. In our experiment, a variable weak magnetic field (B), generated by a pair of rectangular coils surrounding the helium cell, removes the degeneracy of the lower sublevels. These coils are able to produce a constant horizontal magnetic field perpendicular to the direction of propagation of the laser beams. We theoretically estimate the magnetic field, which is constant within the cell area, and experimentally verify it by a teslameter.

\section{Dark resonance profiles in basic tripod configurations}

We consider two different tripod configurations, with the magnetic field parallel or perpendicular to the probe beam polarization. The direction of the static magnetic field is taken as the quantization axis. For the helium $2^3{S_{1}}$ state, the Land\'e $g$-factor is 2.002. The magnetic field shifts the metastable $2^3{S_{1}}$ ($m_{J}$) state by $\mu_B$B$m_J g$, where $\mu_B = e \hbar /2m_e$ = 9.274 $\times$ 10$^{-24}$ J/T is the Bohr magneton, and B is the applied magnetic field. This gives the Zeeman splitting, $\Delta_{\mathrm{Z}} \equiv \mu_B \mathrm{B} m_J g /h$ = 2.8 kHz for B = 1 mG. The Rabi frequency of the coupling beam $\Omega_{\mathrm{C}}$ is much larger than the Zeeman splitting $\Delta_{\mathrm{Z}}$.

In the rotating-wave approximation \cite{Scully97}, the Hamiltonian of the system can be expressed as
\begin{eqnarray}
\textbf{H} = \textbf{H$_0$}+\textbf{H$_I$}.
\end{eqnarray}
\textbf{H$_0$} is the unperturbed Hamiltonian,
\begin{eqnarray}
\textbf{H$_0$} = \displaystyle\sum_{i} \hbar \omega_i \vert i \rangle \langle i \vert ,
\end{eqnarray}
where $i = e, g_-, g_0,g_+$ corresponds to the different levels, labeled in Figs.\,\ref{level145}(b) and \ref{level100}(b). \textbf{H$_I$} is the interaction Hamiltonian involving the coupling and probe transitions.

The time evolution of the density matrix operator, in the presence of decay, is obtained from the Liouville equation as
\begin{equation}
\frac{d}{dt} {\mbox{\boldmath $\rho$}} = -\frac{i}{\hbar} [\textbf{H}, \mbox{\boldmath $\rho$}] + \textbf{R} \mbox{\boldmath $\rho$}, \label{OpticalBlochEq}
\end{equation}
where $\textbf{R}$ is the relaxation matrix. The density matrix elements obey the conditions $\sum_{i} \rho_{ii}=1$ and $\rho_{li} =  \rho^*_{il}$. The sources of relaxation in our system are spontaneous emission from the excited state to the lower states with equal decay rates $\Gamma_0/3$ ($\Gamma_0 = 10^7$ s$^{-1}$), transit relaxation of the atoms through the beams from all allowed states with a rate $\Gamma_{\mathrm{t}}$ ($\approx 10^3$ s$^{-1}$), and Raman coherence decay with a rate $\Gamma_{\mathrm{R}}$ ($\approx$ $10^4$ s$^{-1}$). In our simple model, we do not explicitly take into account the Doppler effect, but assume that the optical coherence decay rate $\Gamma$/2$\pi$ would effectively be given by the width ($\approx$ 1 GHz) of the transition in the Doppler-broadened medium. This approximation has already been shown to be valid in the case of EIT in a standard three-level system in $^4$He* \cite{Goldfarb08,Joyee09}.

\subsection{First configuration} \label{first}

\begin{figure}[htbp]
\begin{center}
 \scalebox{0.45}{\includegraphics{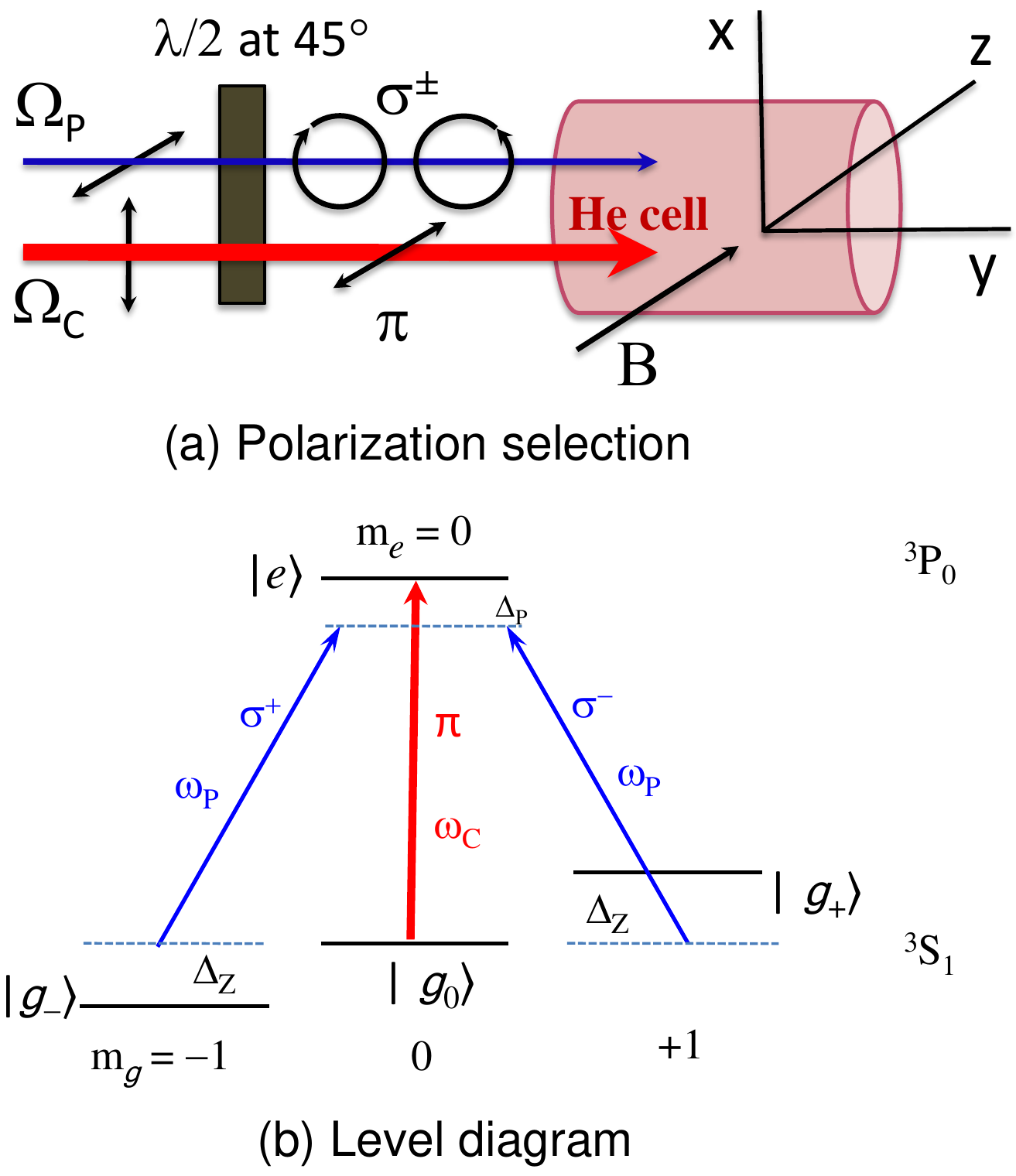}}
\end{center}
\caption{(Color online) Tripod configuration with V-polarized ($\sigma^{\pm}$) probes and H-polarized ($\pi$) coupling.}
    \label{level145}
\end{figure}

When we set the $\lambda/2$ plate in front of our helium cell at 45$^o$ to the incident polarizations, the probe beam has V-polarization ($\sigma$), perpendicular to the magnetic field, while the coupling beam has H-polarization ($\pi$), parallel to the magnetic field (Fig.\,\ref{level145}). Levels $\vert e \rangle$ and $\vert g_- \rangle $($\vert g_+ \rangle $) are coupled by the $\sigma^+$($\sigma^-$)-polarized components of the weak probe beam of frequency $\omega_{\mathrm{P}}$ and detunings $\Delta_\mathrm{P} = \omega_{eg_{\mp}} - \omega_\mathrm{P} \mp \Delta_\mathrm{Z}$. The strong coupling beam of frequency $\omega_{\mathrm{C}}$ and detuning $\Delta_\mathrm{C} = \omega_{eg_0} - \omega_\mathrm{C}$ couples the same excited level $\vert e \rangle$ with the level $\vert g_0 \rangle$. The Raman detuning is $\delta = \Delta_\mathrm{P} - \Delta_\mathrm{C}$.

\begin{figure}[htbp]
\begin{center}
 \scalebox{0.30}{\includegraphics{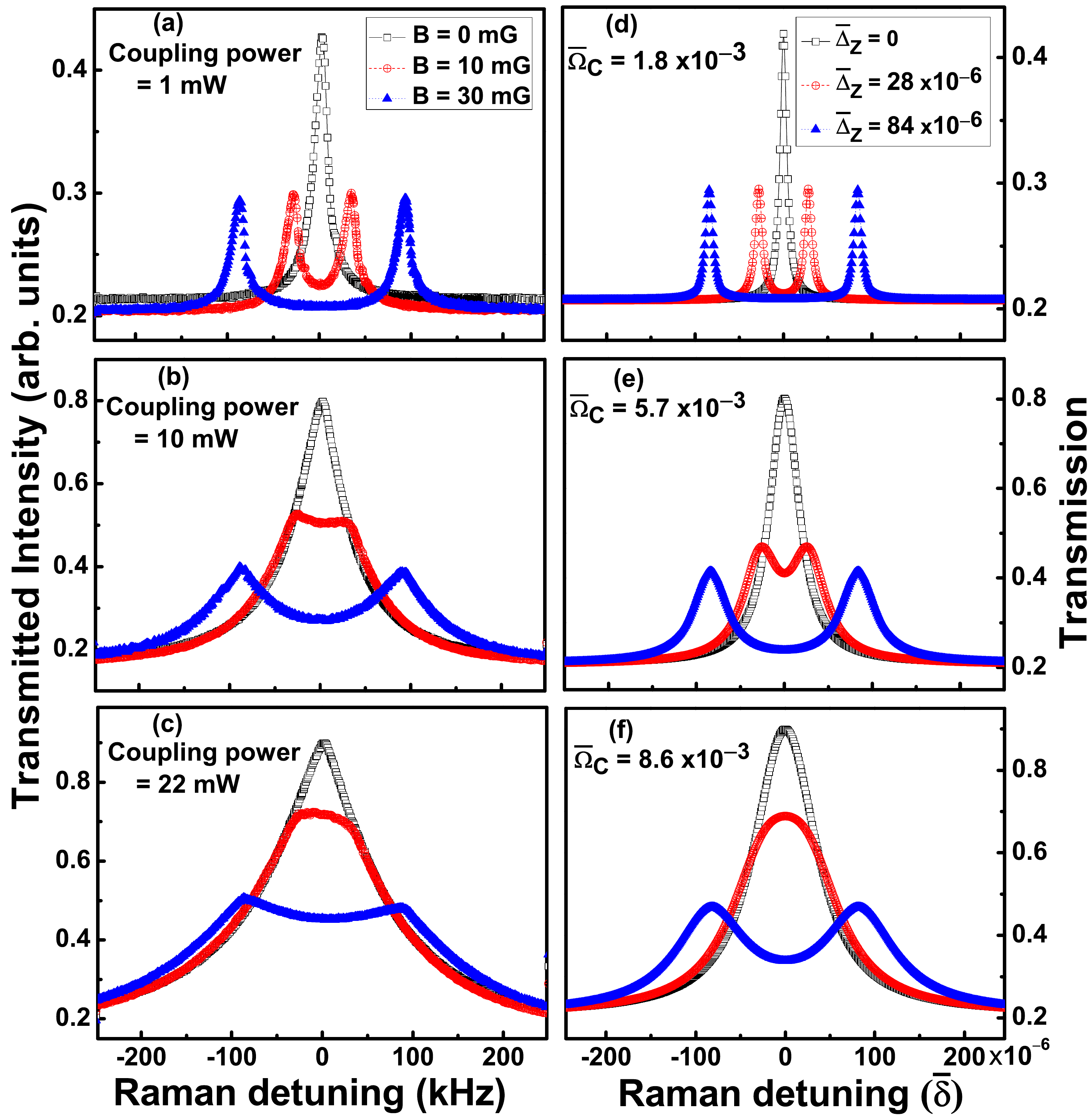}}
\end{center}
\caption{(Color online) Left panel: Experimentally measured transmitted intensity (arb. units) versus Raman detuning $\delta$, corresponding to the configuration shown in Fig.\,\ref{level145}, with magnetic field B at 0 (black, open square), 10 (red, circle) and 30 (blue, triangle) mG, at coupling powers of (a) 1 mW, (b) 10 mW and (c) 22 mW. Right panel, (d), (e) and (f): Corresponding numerically calculated transmission profiles, with $\bar{\Omega}_{\mathrm{C}} \equiv 2 \pi \Omega_{\mathrm{C}}/\Gamma$, $\bar{\delta} \equiv 2 \pi \delta/\Gamma$, and $\bar{\Delta}_{\mathrm{Z}} \equiv 2 \pi \Delta_{\mathrm{Z}}/\Gamma$.}
    \label{Vprobe}
\end{figure}

We experimentally measure the evolution of the transmitted probe intensity (in arbitrary units) versus Raman detuning ($\delta$), as shown in Figs.\,\ref{Vprobe}(a)-(c), for coupling powers of 1 mW, 10 mW and 22 mW, respectively, with magnetic fields of 0, 10 and 30 mG at each coupling power.

We model the system by writing the optical Bloch equations (\ref{OpticalBlochEq}) with the relevant interaction Hamiltonian, for a coupling beam of Rabi frequency $\Omega_{\mathrm{C}}$ with horizontal linear polarization ($\pi$) and a probe beam of Rabi frequency $\Omega_{\mathrm{P}}$ with two counter-circular polarization components ($\sigma^\pm$) with respect to the quantization axis (z):
\begin{eqnarray}
\textbf{H$_I$} = -\frac{\hbar}{2} \Big[ \frac{\Omega_{\mathrm{P}}}{\sqrt{2}} e^{-i \omega_{\mathrm{P}} t} \vert e \rangle \langle g_- \vert + \Omega_{\mathrm{C}} e^{-i \omega_{\mathrm{C}} t} \vert e \rangle \langle g_0 \vert \nonumber \\
+ \frac{\Omega_{\mathrm{P}}}{\sqrt{2}} e^{-i \omega_{\mathrm{P}} t} \vert e \rangle \langle g_+ \vert + \mbox{H.c.} \Big].
\end{eqnarray}
We take $\Omega_{\mathrm{P}}$ $\ll$ $\Omega_{\mathrm{C}}$ and consider $\Omega_{\mathrm{P}}$ to first order. 

We assume that the coupling beam is at resonance ($\Delta_\mathrm{C} = 0$), the populations $\rho_{g_-g_-}$ and $\rho_{g_+g_+}$ are approximately equal to 0.50, and $\rho_{g_0 g_0} \approx 0 \approx \rho_{ee}$. Then the steady state solutions of the six coupled optical Bloch equations for the coherences $\rho_{eg_-} = \tilde{\rho}_{eg_-} e^{-i \omega_{\mathrm{P}}t}$, $\rho_{eg_+} = \tilde{\rho}_{eg_+} e^{-i \omega_{\mathrm{P}}t}$, $\rho_{g_0 g_-} = \tilde{\rho}_{g_0 g_-} e^{i (\omega_{\mathrm{C}} - \omega_{\mathrm{P}})t}$, $\rho_{g_0 g_+} = \tilde{\rho}_{g_0 g_+} e^{i (\omega_{\mathrm{C}} - \omega_{\mathrm{P}})t}$, $\rho_{g_+ g-} = \tilde{\rho}_{g_+g_-}$ and $\rho_{g_- g_+} = \tilde{\rho}_{g_- g_+}$ give 
\begin{eqnarray}
 \frac{\tilde{\rho}_{eg_\mp} } {\bar{\Omega}_{\mathrm{P}}}&=&
 \frac{w_{\mp} /\sqrt{2}  } {2 ( a_{\mp} + \bar{\Delta}_\mathrm{C} - \frac{i}{3}) - \frac{\vert \bar{\Omega}_{\mathrm{C}} \vert ^2 } {2(  a_{\mp} - i \bar{\Gamma}_{\mathrm{R}})}},
\end{eqnarray}
where $w_{\mp} = ( \rho_{g_\mp g_\mp} - \rho_{ee}) = 0.5$, $a_{\mp} = \bar{\delta} \mp \bar{\Delta}_{\mathrm{Z}}$, and all rates and frequencies, scaled by $\Gamma/2\pi \approx 10^9$ Hz, are denoted by a bar over the corresponding symbols. 

The probe absorption and dispersion are proportional to the imaginary and real parts of the susceptibility. We obtain an expression for the probe susceptibility as
\begin{eqnarray}
\chi (\omega_{\mathrm{P}}) = \frac{A_{1}}{2 \sqrt{2}} \left[\frac{1} { (a_{+} - \frac{i}{3}) - \frac{\vert \bar{\Omega}_{\mathrm{C}} \vert ^2 } {4 (a_{+} - i \bar{\Gamma}_{\mathrm{R}})}}\right. \nonumber \\
\left. + \frac{1 } {(a_{-} - \frac{i}{3}) - \frac{\vert \bar{\Omega}_{\mathrm{C}} \vert ^2 } {4 (a_{-} - i \bar{\Gamma}_{\mathrm{R}})}}\right] ,
 \label{eq-chiP}
\end{eqnarray}
where $A_{1} = \mathcal{N} \vert \mu_{eg} \vert ^2 w_{\mp}/\hbar \epsilon_0$, $\mu_{eg_-} \approx \mu_{eg_+} = \mu_{eg}$ is the dipole matrix element for the probe transitions, and $\mathcal{N}$ is the atomic density. With the above reasonable approximations, the imaginary part of the susceptibility from Eq.\,(\ref{eq-chiP}) is found to be
\begin{eqnarray}
\mathrm{Im} \left[ \chi (\omega_{\mathrm{P}})  \right] = \frac{3 A_{1}}{\sqrt{2}}
 \left[1-\frac{3 \vert \bar{\Omega}_{\mathrm{C}} \vert ^2}{8}  \left(\frac{\Lambda }{a^2_{+}
 + \Lambda ^2} \right.\right. \nonumber \\
\left.\left. + \frac{\Lambda }{a^2_{-} + \Lambda ^2}\right) \right],
   \label{Lorentzian2}
\end{eqnarray}
where $\Lambda = \bar{\Gamma}_{\mathrm{R}} + \frac{3 \vert \bar{\Omega}_{\mathrm{C}} \vert ^2}{4}$. The right-hand side of (\ref{Lorentzian2}) is a sum of two Lorentzians with centers at $\bar{\delta}$= $\pm \bar{\Delta}_\mathrm{Z}$ and full widths at half maxima of $2 \Lambda$. The transmission profiles are generated from $\mathrm{exp}\left[ -k \mathrm{L} \mathrm{Im} \left[ \chi (\omega_{\mathrm{P}}) \right] \right]$, where $k$ is the magnitude of the wave vector of the probe beam, and L is the length of the helium cell. The transmission profiles versus scaled Raman detuning ($\bar{\delta}$) are shown in Figs.\,\ref{Vprobe}(d)-(f), with $\bar{\Omega}_{\mathrm{C}}$ = 1.8 $\times$ 10$^{-3}$, 5.7 $\times$ 10$^{-3}$ and 8.6 $\times$ 10$^{-3}$, respectively, corresponding to the experimental coupling powers, with Zeeman shifts also corresponding to the experimental values of the magnetic field.

In this configuration, as seen in Fig.\,\ref{Vprobe}, at zero magnetic field, we observe a single EIT peak at the line center \cite{Lezama99}. When we apply a weak magnetic field, we observe double dark resonances for low coupling powers. The two corresponding peaks add incoherently. As we increase the coupling power, these two peaks broaden and eventually merge together into a single peak at the line center. To observe double dark resonances, it is required that all population is optically pumped into the $\vert g_- \rangle$ and $\vert g_+ \rangle$ levels. The fact that this configuration leads to an incoherent sum of two EIT peaks can be easily understood. Indeed, the weak probe (treated to first order in the Rabi frequency $\Omega_{\mathrm{P}}$) cannot create any coherence between the two populated probe ground levels, $\vert g_- \rangle$ and $\vert g_+ \rangle$. We can thus expect the system to behave as two independent three-level systems connected by the single coupling beam, with each three-level system exhibiting its respective EIT peak for its particular Raman resonance. 

\subsection{Second configuration} \label{second}

\begin{figure}[htbp]
\begin{center}
 \scalebox{0.45}{\includegraphics{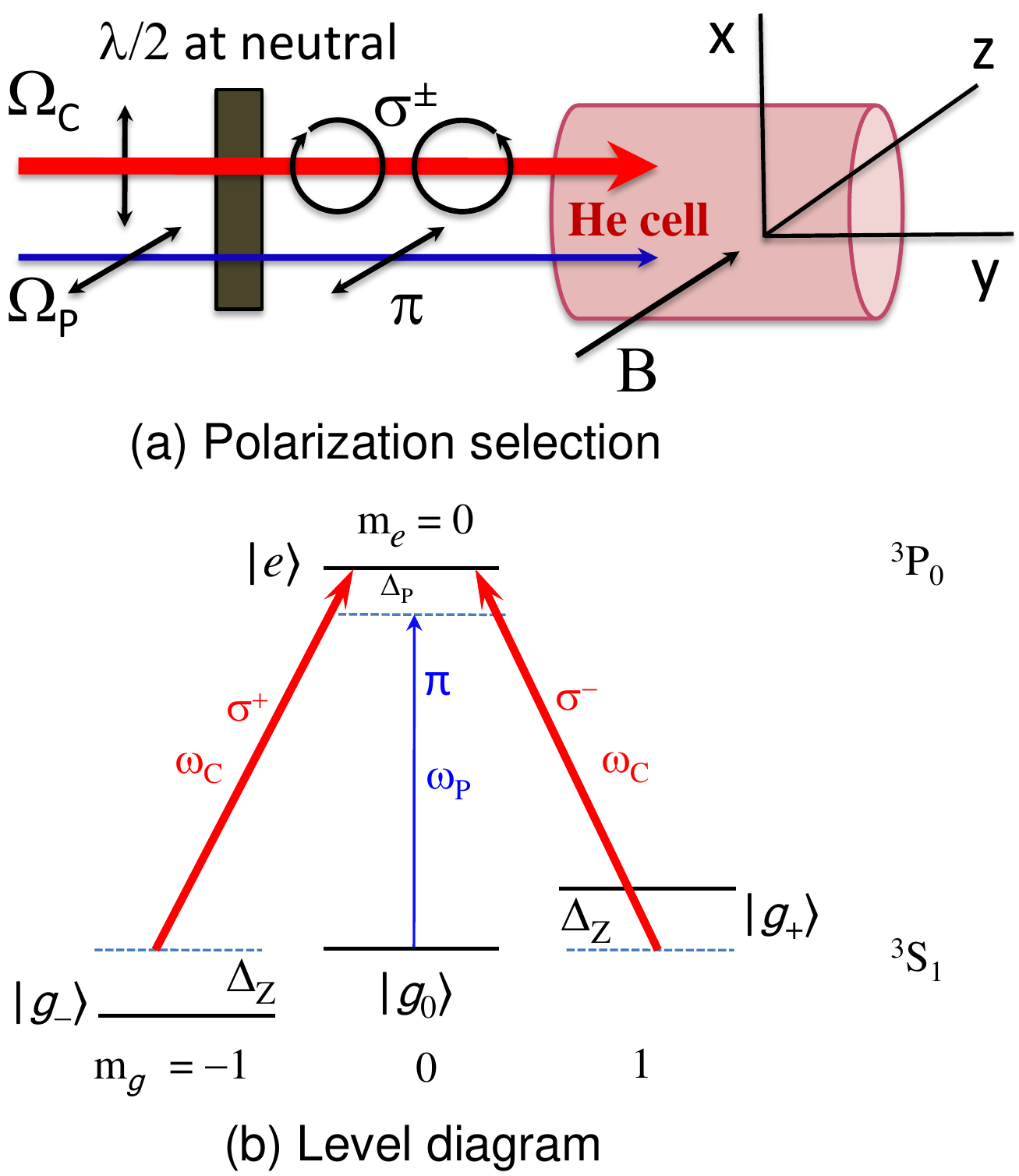}}
\end{center}
\caption{(Color online) Tripod configuration with H-polarized ($\pi$) probe and V-polarized ($\sigma^{\pm}$) coupling beams.}
\label{level100}
\end{figure}

We now set the $\lambda/2$ plate in front of the helium cell at a specific angle so that this plate behaves as neutral for the incident polarizations -- the probe beam has H-polarization ($\pi$), parallel to the magnetic field, while the coupling beam has V-polarization ($\sigma$), perpendicular to the magnetic field (see Fig.\,\ref{level100}). Levels $\vert e \rangle$ and $\vert g_{-} \rangle $($\vert g_{+} \rangle $) are coupled by the $\sigma^+$($\sigma^-$)-polarized component of the strong coupling beam of frequency $\omega_\mathrm{C}$ and detunings $\Delta_\mathrm{C}$ = $\omega_{eg_{\mp}} - \omega_\mathrm{C} \mp \Delta_\mathrm{Z}$. A weak probe beam of frequency $\omega_\mathrm{P}$ and detuning $\Delta_\mathrm{P}$ = $\omega_{eg_{0}} - \omega_\mathrm{P}$ couples the same excited level $\vert e \rangle$ with the level $\vert g_{0} \rangle$.

We again model the system by writing the density matrix equations (\ref{OpticalBlochEq}), with the corresponding interaction Hamiltonian for a probe beam of Rabi frequency $\Omega_{\mathrm{P}}$ with horizontal linear polarization ($\pi$) and a coupling beam of Rabi frequency $\Omega_{\mathrm{C}}$ with two counter-circular polarization components ($\sigma^\pm$) with respect to the quantization axis:
\begin{eqnarray}
\textbf{H$_I$} = -\frac{\hbar}{2} \Big[\frac{\Omega_{\mathrm{C}}}{\sqrt{2}} e^{-i \omega_{\mathrm{C}} t} \vert e \rangle \langle g_- \vert + \Omega_{\mathrm{P}} e^{-i \omega_{\mathrm{P}} t} \vert e \rangle \langle g_0 \vert \nonumber \\
+ \frac{\Omega_{\mathrm{C}}}{\sqrt{2}} e^{-i \omega_{\mathrm{C}} t} \vert e \rangle \langle g_+ \vert + \mbox{H.c.} \Big].
\end{eqnarray}
The main approximations used are to take $\Omega_\mathrm{P}$ $\ll$ $\Omega_\mathrm{C}$, and to consider $\Omega_\mathrm{P}$ to first order while taking $\Omega_\mathrm{C}$ in all orders.

In our case, $\Delta_\mathrm{C} = 0$, as before. We assume that the probe ground level population  $\rho_{g_0g_0}$ is approximately equal to unity, and $\rho_{g_- g_-} \approx 0 \approx \rho_{g_+ g_+} \approx \rho_{ee}$. Then the steady state solutions of the three coupled optical Bloch equations for the coherences $\rho_{eg_0} = \tilde{\rho}_{eg_0} e^{-i \omega_{\mathrm{P}}t}$, $\rho_{g_- g_0} = \tilde{\rho}_{g_- g_0} e^{-i (\omega_{\mathrm{P}}-\omega_{\mathrm{C}})t}$ and $\rho_{g_+ g_0} = \tilde{\rho}_{g_+ g_0} e^{-i (\omega_{\mathrm{P}} - \omega_{\mathrm{C}})t}$ give
\begin{eqnarray}
\frac{\tilde{\rho}_{eg_0}} {\bar{\Omega}_{\mathrm{P}}} &=& \frac{w_{0}} {2 b +\frac{\vert \bar{\Omega}_\mathrm{C} \vert ^2} {4} \Big[\frac{1} {( i \bar{\Gamma}_{\mathrm{R}} - a_{-})}  - \frac{1} {(a_{+} - i  \bar{\Gamma}_{\mathrm{R}})}\Big]},
\end{eqnarray}
where $w_{0} = ( \rho_{g_0 g_0} - \rho_{ee} )$ = 1, $a_{\mp} = \bar{\delta} \mp \bar{\Delta}_{\mathrm{Z}}$, $b = \bar{\delta} + \bar{\Delta}_{\mathrm{C}} - \frac{i}{3}$, and all rates and frequencies, scaled by $\Gamma/2\pi \approx 10^9$ Hz, are denoted by a bar on top, as before. 

We obtain an expression for the probe susceptibility as
\begin{eqnarray}
\chi (\omega_{\mathrm{P}}) &=& \frac{A_{2} (a_{-} - i  \bar{\Gamma}_{\mathrm{R}})(a_{+} - i  \bar{\Gamma}_{\mathrm{R}})} {2 b (a_{-} - i \bar{\Gamma}_{\mathrm{R}})(a_{+} - i  \bar{\Gamma}_{\mathrm{R}}) - q\frac{\vert \bar{\Omega}_\mathrm{C} \vert ^2 }{2 }},
  \label{eq-chiP1}
\end{eqnarray}
where $q = ( \bar{\delta} - i \bar{\Gamma}_{\mathrm{R}})$, $A_{2} = \mathcal{N} \vert \mu_{eg_0} \vert ^2 w_{0}/\hbar \epsilon_0$, and $\mu_{eg_0}$ is the dipole matrix element for the probe transition. The absorption and dispersion of the probe beam are proportional to the imaginary and real parts of the susceptibility. With the above reasonable approximations, we get the imaginary part of the susceptibility from Eq.\,(\ref{eq-chiP1}) as
\begin{equation}
\mathrm{Im} \left[ \chi (\omega_{\mathrm{P}})  \right] = 3 A_{2} \frac{2 a^{2}_{-} a^{2}_{+} +  \bar{\Gamma}_{\mathrm{R}} y (2 \bar{\Gamma}_{\mathrm{R}}+x)+ \bar{\Gamma}^{3}_{\mathrm{R}}x}{4 a^{2}_{-} a^{2}_{+} +  4 \bar{\Gamma}_{\mathrm{R}} x y+x^2\bar{\Gamma}^{2}_{\mathrm{R}}+9 \bar{\delta}^{2} \frac{\vert \bar{\Omega}_{\mathrm{C}} \vert ^4}{4} },
\label{Lorentzian1}
\end{equation}
where $x = 2 \bar{\Gamma}_{\mathrm{R}}+3 \frac{\vert \bar{\Omega}_{\mathrm{C}} \vert ^2}{2}$ and $y = \bar{\delta}^2 + \bar{\Delta}_{\mathrm{Z}}^2$. 

\begin{figure}[htbp]
\begin{center}
 \scalebox{0.30}{\includegraphics{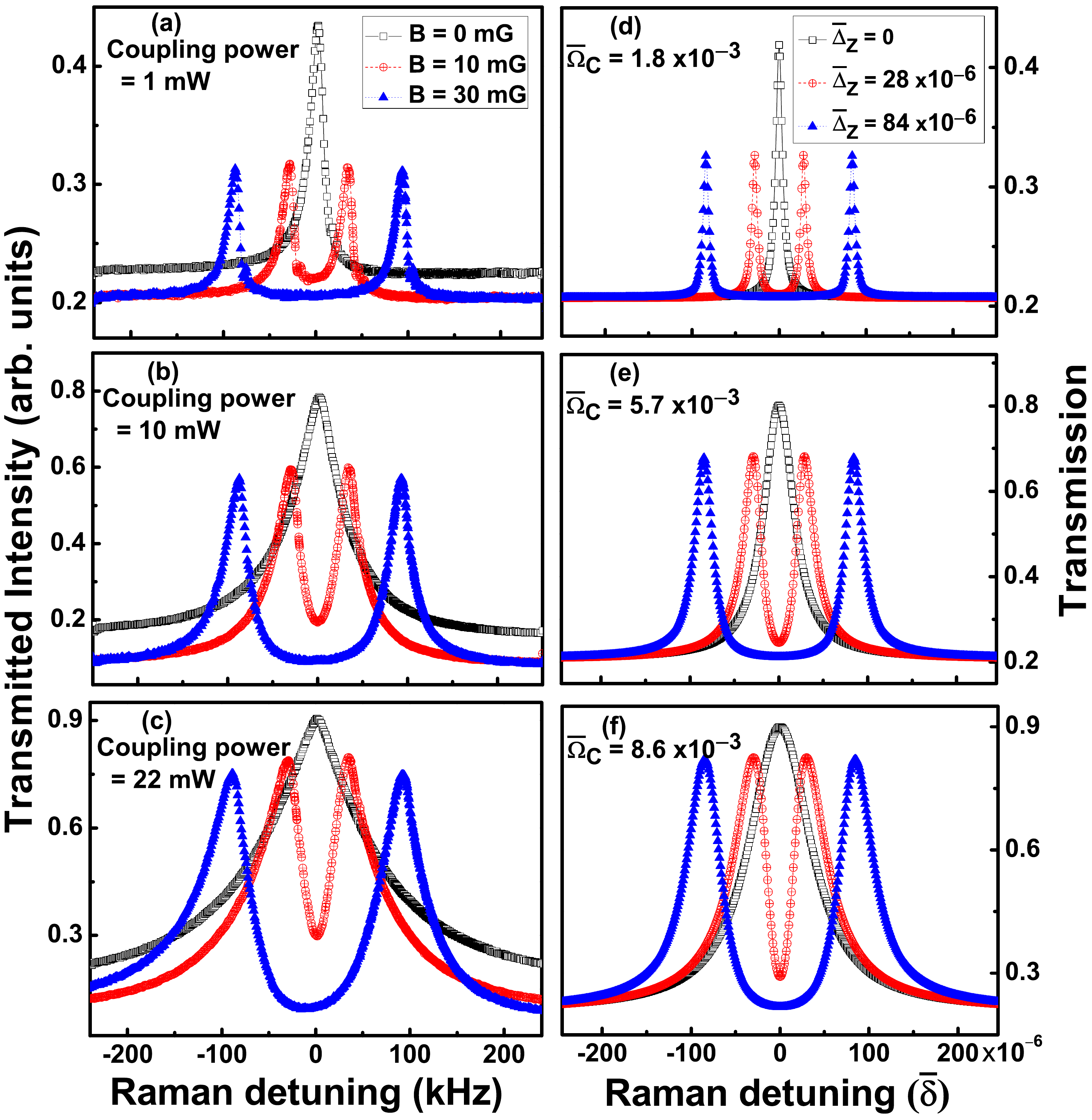}}
\end{center}
\caption{(Color online) Left panel: Experimentally measured transmitted intensity (arb. units) versus Raman detuning $\delta$, corresponding to the configuration shown in Fig.\,\ref{level100}, with magnetic field B at 0 (black, open square), 10 (red, circle) and 30 (blue, triangle) mG for coupling powers of (a) 1 mW, (b) 10 mW, and (c) 22 mW. Right panel, (d), (e) and (f): Corresponding numerically calculated transmission profiles, with $\bar{\Omega}_{\mathrm{C}} \equiv 2 \pi \Omega_{\mathrm{C}}/\Gamma$, $\bar{\delta} \equiv 2 \pi \delta/\Gamma$, and $\bar{\Delta}_{\mathrm{Z}} \equiv 2 \pi \Delta_{\mathrm{Z}}/\Gamma$.}
    \label{Hprobe}
\end{figure}

We plot the experimentally measured transmitted intensity (arb. units) in Figs.\,\ref{Hprobe}(a)-(c), for coupling powers of 1 mW, 10 mW and 22 mW. The corresponding numerically calculated transmission profiles versus scaled Raman detuning are reproduced in Figs.\,\ref{Hprobe}(d)-(f), with the corresponding values of $\bar{\Omega}_{\mathrm{C}}$ = 1.8 $\times$ 10$^{-3}$, 5.7 $\times$ 10$^{-3}$ and 8.6 $\times$ 10$^{-3}$, respectively. In each case, we plot the profiles matching the magnetic fields of 0, 10 and 30 mG, as in the experiment. Our numerical simulations are in good agreement with the experimental results.

This configuration, at zero magnetic field, as shown in Fig.\,\ref{Hprobe}, is equivalent to the degenerate two-level system with a $\sigma^{\pm}$ coupling and a $\pi$ probe. In this case, we observe a single EIT peak (black, open square) at the line center \cite{Goldfarb08}. Note that Kim {\it et al.} \cite{Kim03} observed electromagnetically induced absorption (EIA) for the two-level degenerate system ($F_{e} = F_{g}-1$ with $F_{e} \geq$ 1 and $F_{e} = F_{g}$) and this anomalous EIA has been interpreted by the analysis of dressed-atom multiphoton spectroscopy \cite{Chou10}. In our degenerate two-level system ($F_{e} = F_{g}-1$ with $F_{e} = 0$), in place of anomalous EIA, we observe an EIT peak at the line center. This EIT peak can be explained by the following change of basis, which is a simple example of Morris-Shore transformation \cite{Kyoseva06}: Replace the two sublevels $\vert g_- \rangle$ and $\vert g_+ \rangle$ by the usual dark ($\vert NC \rangle = (\vert g_- \rangle - \vert g_+ \rangle)/\sqrt{2}$) and bright ($\vert C \rangle = (\vert g_- \rangle + \vert g_+ \rangle)/\sqrt{2}$) states for the CPT in a three-level system ($\vert g_-\rangle$, $\vert g_+ \rangle$ and $\vert e \rangle$) for the coupling beams. Since the transition $\vert NC\rangle \rightarrow \vert e \rangle$ is not allowed, we have essentially obtained a three-level system ($\vert C \rangle$, $\vert g_0 \rangle$ and $\vert e \rangle$) with a probe and a coupling beam. Hence we observe a single EIT peak at the line center. The widths and heights of the EIT windows increase with the coupling power. 

When we apply a weak transverse magnetic field, the degeneracy of the lower levels is removed. In this case, two dark resonance peaks appear and they shift from the zero detuning position with increasing magnetic field. The separation between the two dark resonances varies linearly with the applied magnetic field. The double dark resonance cannot be explained in terms of transfer of coherence from the excited level to the ground level \cite{Goren04,Meshulam07} but these dark resonances are two EIT peaks at $\delta = \pm \Delta_{Z}$. When the coupling power is increased in the presence of the magnetic field, an absorption line appears, much narrower and deeper than that found in the first configuration, which is the signature of an interference phenomenon between the two induced EIT windows \cite{Yelin03,Zhang07}. This is the main result of this paper. It is visible in both the experimental data and in simulations that these EIT peaks are asymmetric at 10 and 22 mW coupling powers: the absorption dip walls are sharper than the external transparency window lines. Although it is well known that EIT profiles become asymmetric when the coupling beam is optically detuned, the optical detuning given by the Zeeman shift cannot explain such a shape: indeed, the detuning in our case is less than 100 kHz while hundreds of MHz are necessary to obtain any significant asymmetry in our system \cite{Goldfarb09}. When comparing these profiles with the ones obtained with the first configuration (see Fig.\,\ref{Vprobe}), one notices here that the double peak transmissions are much higher. For the two values of the Zeeman shift used and at high enough coupling powers (10 and 22 mW), the transmissions are nearly as large as the transmission of the single peak recorded without any magnetic field: transmissions seem to be given by the total Rabi frequency $\Omega_\mathrm{C}$ while the widths are much narrower than the width expected with such a coupling intensity. The resulting very deep absorption line seems to narrow with increasing coupling power instead of disappearing because of saturation. This is very different from the first case, where the transmissions corresponding to 80 kHz of Zeeman shift remain roughly half the transmission of this single peak, and the saturation broadening makes the dip disappear for 50 kHz of Zeeman shift. We have checked that an incoherent addition of susceptibilities or transmissions would give a behavior similar to the first case and cannot explain the data recorded in the second configuration: the absorption dip appearing at the line center is narrower and deeper than is possible by adding two independent best-fit EIT profiles. A common picture for EIT in three-level systems is to consider that it is the result of interference between two absorption paths, a direct absorption from the probed level, and the other which is followed by induced emission and reabsorption by the coupling beam. In our tripod configuration, there can be emission and reabsorption with both $\sigma^+$ and $\sigma^-$coupling beams. The constructive or destructive nature of this interference mechanism depends on the signs of the superpositions in the two dark states corresponding to the two three-level systems. In our case, the components of the coupling beam lead to opposite signs for the $|g_- \rangle \leftrightarrow |e \rangle$ and $|g_+ \rangle \leftrightarrow |e \rangle$ transition amplitudes. As a result, there is a destructive interference between two EIT peaks at the center and we observe a sharp absorption dip, looking like (but different in nature from) EIA \cite{Taichenachev99}, flanked by two EIT (detuned) peaks \cite{Goren04}. It is clear from Fig.\,\ref{Hprobe} that for fixed magnetic fields, the widths and heights of all the EIT windows increase while the width of the absorption dip decreases with increasing coupling power (the full-widths at half-maxima of the absorption dips are about 47, 42 and 34 kHz, for coupling powers of 1, 10 and 22 mW, respectively, at B = 10 mG). In the absence of the magnetic field, the narrow absorption dip disappears and the system becomes transparent at the line center.

\section{Conclusions}

We have been able to carve out a clean four-level tripod system in a simple system of room-temperature $^4$He* using a weak magnetic field and polarization-selective transitions. Interesting interplay between the double dark resonances has been recorded.

In the first tripod configuration with two probed transitions, when the coupling is parallel to the weak magnetic field, we have observed that two EIT (detuned) add incoherently, and for large coupling power, these two peaks merge into a single EIT like peak at the line center. Such a double-EIT configuration has potential application in light storage for two frequencies \cite{Karpa08,Petrosyan04} and coupling-induced switch in the presence of a small magnetic field.

In the second tripod configuration with two coupling transitions, when the probe is parallel to the weak magnetic field, we have observed a remarkable destructive interference between the two EIT peaks, leading to an extra, narrow absorption peak in-between the EIT peaks. The absorption feature is seen to become narrower with increasing coupling power and could be made subnatural even in the presence of Doppler broadening \cite{Goren04}. We stress here that our results, shown in Fig.\,\ref{Hprobe}, cannot be obtained from two independent EIT systems, even if one allows for asymmetric EIT windows resulting from detuned (by a few tens of kHz) coupling fields. The absorption dip appearing at the line center is narrower and deeper than is possible by adding two independent asymmetric EIT fits. The separation between the observed EIT peaks is determined by the applied magnetic field. Such a system may be used as a magnetometer, although the field values used by us are rather high. The shape of the resonances, however, depends critically on the direction of the magnetic field, making the sensor anisotropic. For known directions of the magnetic field, the symmetry of the system offers a specific advantage based on measurements of {\it differences} in frequencies \cite{Yudin10}. In the absence of the magnetic field, the narrow absorption maximum disappears and the system becomes transparent at the line center. It thus has the potential to be used as a magneto-optic switch, with pulsed operation. 

We have successfully modeled the system. For mixed polarizations of the coupling and the probe along the quantization axis, the structure of the resonances becomes complex and the features are under further investigation.

\begin{acknowledgments}

This work is supported by the Indo-French Centre for Advanced Research (IFCPAR/CEFIPRA). The work of SK is supported by the Council of Scientific and Industrial Research, India.

\end{acknowledgments}


\begin{thebibliography}{99}

\bibitem{Harris97} S.E. Harris, Phys. Today {\bf 50}, 36 (1997), and references therein.

\bibitem{Lukin99} M.D. Lukin, S.F. Yelin, M. Fleischhauer, and M.O. Scully,  Phys. Rev. A {\bf 60}, 3225 (1999).

\bibitem{Yelin03} S.F. Yelin, V.A. Sautenkov, M.M. Kash, G.R. Welch, and M.D. Lukin, Phys. Rev. A {\bf 68}, 063801 (2003).

\bibitem{Liang05} Y. Niu, S. Gong, R. Li, Z. Xu, and X. Liang, Opt. Lett. {\bf 30}, 3371 (2005).

\bibitem{Mahmoudi08}  M. Mahmoudi, R. Fleischhaker, M. Sahrai, and J. Evers, J. Phys. B: At. Mol. Opt. Phys. {\bf 41}, 025504 (2008).

\bibitem{Bergmann96} J. Martin, B.W. Shore, and K. Bergmann, Phys. Rev. A {\bf 54}, 1556 (1996).

\bibitem{Bergmann03} F. Vewinger, M. Heinz, R.G. Fernandez, N.V. Vitanov, and K. Bergmann, Phys. Rev. Lett. {\bf 91}, 213001 (2003).

\bibitem{Unanyan04} R.G. Unanyan, M.E. Pietrzyk, B.W. Shore, and K. Bergmann, Phys. Rev. A {\bf 70}, 053404 (2004).


\bibitem{Ham00} B.S. Ham and P.R. Hemmer, Phys. Rev. Lett. {\bf 84}, 4080 (2000).

\bibitem{Peng08}  Y. Han, J. Xiao, Y. Liu, C. Zhang, H. Wang, M. Xiao, and K. Peng, Phys. Rev. A {\bf 77}, 023824 (2008).

\bibitem{Sun08} Y. Guo, L. Zhou, L.M. Kuang, and C.P. Sun, Phys.\,Rev.\,A {\bf 78}, 013833 (2008).

\bibitem{Karpa08} L. Karpa, F. Vewinger, and M. Weitz, Phys. Rev. Lett. {\bf 101}, 170406 (2008).

\bibitem{Petrosyan04} D. Petrosyan and Y.P. Malakyan, Phys.\,Rev.\,A {\bf 70}, 023822 (2004).

\bibitem{Lvovsky08} A. MacRae, G. Campbell, and A.I. Lvovsky, Opt. Lett. {\bf 33}, 2659 (2008).

\bibitem{Drampyan09} R. Drampyan, S. Pustelny, and W. Gawlik, Phys.\,Rev.\,A {\bf 80}, 033815 (2009).

\bibitem{Goren04} C. Goren, A.D. Wilson-Gordon, M. Rosenbluh, and H. Friedmann, Phys.\,Rev.\,A {\bf{69}}, 063802 (2004).

\bibitem{Gavra07} N. Gavra, M. Rosenbluh, T. Zigdon, A.D. Wilson-Gordon, and H. Friedmann, Opt. Comm. {\bf 280}, 374 (2007).

\bibitem{Zhang07} J.Q. Shen and P. Zhang, Opt.\,Exp. {\bf 15}, 6484 (2007).

\bibitem{Goldfarb08} F. Goldfarb, J. Ghosh, M. David, J. Ruggiero, T. Chaneli\`ere, J.-L. Le Gou\"et, H. Gilles, R. Ghosh, and 
F. Bretenaker, Europhys. Lett. {\bf 82}, 54002 (2008).

\bibitem{Joyee09} J. Ghosh, R. Ghosh, F. Goldfarb, J.-L. Le Gou\"et, and F. Bretenaker, Phys. Rev. A {\bf 80}, 023817 (2009).

\bibitem{Shlyapnikov94} G.V. Shlyapnikov, J.T.M. Walraven, U.M. Rahmanov, and M.W. Reynolds, Phys. Rev. Lett. {\bf 73}, 3247 (1994).

\bibitem{Taichenachev99} A.V. Taichenachev, A.M. Tumaikin, and V.I. Yudin, Phys.\,Rev.\,A {\bf{61}}, 011802(R) (1999).

\bibitem{Meshulam07} R. Meshulam, T. Zigdon, A.D. Wilson-Gordon, and H. Friedmann, Opt. Lett. {\bf 32}, 2318 (2007).

\bibitem{Scully97} M.O. Scully and M.S. Zubairy, \textit{Quantum Optics} (Cambridge University Press, Cambridge, 1997).

\bibitem{Lezama99} A. Lezama, S. Barreiro, A. Lipsich, and A.M. Akulshin, Phys.\,Rev.\,A {\bf{61}}, 013801 (1999).

\bibitem{Kim03} S.K. Kim, H.S. Moon, K. Kim, and J.B. Kim, Phys.\,Rev.\,A {\bf{68}}, 063813 (2003).

\bibitem{Chou10} H.S. Chou and J. Evers, Phys.\,Rev.\,Lett. {\bf{104}}, 213602 (2010).

\bibitem{Kyoseva06} E.S. Kyoseva and N.V. Vitanov, Phys.\,Rev.\,A \textbf{73}, 023420 (2006).

\bibitem{Goldfarb09} F. Goldfarb, T. Laupr\^etre, J. Ruggiero, F. Bretenaker, J. Ghosh and R. Ghosh, C.R. Physique {\bf{10}}, 919 (2009).

\bibitem{Yudin10} V.I. Yudin, A.V. Taichenachev, Y.O. Dudin, V.L. Velichansky, A.S. Zibrov, and S.A. Zibrov, Phys.\,Rev.\,A {\bf{82}}, 033807 (2010).

\end{thebibliography}
\end{document}